\documentclass[conference]{IEEEtran}
\IEEEoverridecommandlockouts
\usepackage{cite}
\usepackage{amsmath,amssymb,amsfonts}
\usepackage{algorithmic}
\usepackage{mathtools}
\usepackage{graphicx}
\usepackage{textcomp}
\usepackage{caption,subcaption}
\usepackage{xfrac}
\usepackage{color}
\usepackage{float}
\usepackage[normalem]{ulem}
\usepackage[linesnumbered,ruled,norelsize]{algorithm2e}
\usepackage{listings}
\definecolor{dkgreen}{rgb}{0,0.6,0}
\definecolor{gray}{rgb}{0.5,0.5,0.5}
\definecolor{mauve}{rgb}{0.58,0,0.82}
\usepackage{xcolor}
\def\BibTeX{{\rm B\kern-.05em{\sc i\kern-.025em b}\kern-.08em
    T\kern-.1667em\lower.7ex\hbox{E}\kern-.125emX}}

\newcommand{\mg}[1]{{\color{black} #1}}

\newcommand{\framework}{ASYNC}

    \usepackage{hyperref}

\begin{document}

\title{ASYNC: A Cloud Engine with Asynchrony and History for Distributed Machine Learning\\
}
\author{\IEEEauthorblockN{Saeed Soori\IEEEauthorrefmark{1},
Bugra Can\IEEEauthorrefmark{2}, Mert Gurbuzbalaban\IEEEauthorrefmark{3} and
Maryam Mehri Dehnavi\IEEEauthorrefmark{4}}
\IEEEauthorblockA{\IEEEauthorrefmark{1} \IEEEauthorrefmark{4} Department of Computer Science,
University of Toronto,
Toronto, Canada\\
\IEEEauthorrefmark{2} \IEEEauthorrefmark{3} Department of MBS,
Rutgers Univeristy,
New Jersey, USA\\
Email: \IEEEauthorrefmark{1}sasoori@cs.toronto.edu,
\IEEEauthorrefmark{2}bugra.can@rutgers.edu,
\IEEEauthorrefmark{3}mert.gurbuzbalaban@rutgers.edu,
\IEEEauthorrefmark{4}mmehride@cs.toronto.edu}}



\maketitle

\begin{abstract}
ASYNC is a  framework that supports the implementation of asynchrony and history for optimization methods on distributed computing platforms. The popularity of asynchronous optimization methods has increased in distributed machine learning. However, their applicability and practical experimentation on distributed systems are limited because current bulk-processing cloud engines do not provide a robust support for asynchrony and history.  With introducing three main modules and bookkeeping system-specific and application parameters, ASYNC provides practitioners with a framework  to implement asynchronous machine learning methods. To demonstrate ease-of-implementation in ASYNC, the synchronous and asynchronous variants of two well-known optimization methods, stochastic gradient descent and SAGA, are demonstrated in \framework{}. 
\end{abstract}


\section{Introduction}

Distributed optimization has gained significant traction in recent years and is frequently used to solve modern large-scale machine learning problems\cite{bottou2012stochastic}. The challenges of dealing with
huge datasets, has lead to the development of optimizations methods with  \textit{asynchrony} and \textit{history}.  Asynchronous optimization methods reduce worker idle times and mitigate communication costs. Operations on a history of gradients augments the noise (stochasticity)  to improve convergence \cite{reddi2015variance}. Distributed optimization methods operate on batches of data and thus have to be implemented in  cluster-computing engines with a bulk (coarse-grained)  computation model. 

There exists several general coarse-grained distributed data processing systems.  Hadoop\cite{hadoop2011apache} and Spark \cite{zaharia2012resilient} are based on the iterative map-reduce model  but use a synchronous iterative communication pattern. Thus, because of not supporting  asynchrony,  their execution is vulnerable to the diverse performance profile caused by slow workers, i.e. stragglers, and network latency in a distributed platform. Also history can not be efficiently maintained in these engines as it requires storing \textit{bulky worker-results}, and introduces overheads to their lineage-based \cite{zaharia2012resilient} or checkpointing fault tolerance implementations.

Recently, a number of coarse-grained machine learning engines such as Petuum\cite{xing2015petuum} and Litz\cite{qiao2018litz}, have adopted the parameter server\cite{jagerman2016web} architecture  to implement asynchronous communication between nodes with  push-pull operations. Asynchrony in distributed optimization methods is implemented with consistency models, i.e. barriers, expressed via a dependency graph that maintain a trade-off between system efficiency and algorithm convergence. Parameter server paradigms implement a specific class of consistency models, i.e. stale synchronous parallel (SSP) paradigms using a fixed dependency graph,  which use a static staleness threshold to control worker wait times. However, recent advancements in distributed optimization \cite{zhang2018adaptive,albrecht2006loose} demand for wider range of customized consistency models (CCMs), often defined by the user  such as  throttled-release\cite{albrecht2006loose}, that control worker wait times using parameters such as worker-task-completion time \cite{zhang2018adaptive} and require to adaptively adjust the parameters at runtime. CCMs can not be implemented in available parameter server frameworks as it needs the underlying dependency graph to adaptively be reconfigured at runtime.  Also,  Petuum does not support history and  Litz preserves the history by periodic checkpointing with significant overheads. 
Other distributed parameter-server frameworks such as DistBelief\cite{dean2012large} and TensorFlow  \cite{abadi2016tensorflow}  are specialized for deep learning applications and thus do not naturally support consistency models and history.


Amongst the fine-grained distributed data processing systems, primarily used for streaming applications, RAY \cite{moritz2018ray} and Flink \cite{carbone2015apache} support asynchronous function invocations with dynamic data flow graphs \cite{wang2019lineage}. However, these frameworks do not support CCMs and are primarily designed for fine-grained tasks, and thus can not naturally extend to a bulk-processing engine. Also, while streaming engines (because of processing fine tasks) can store local results and intermediate data on workers  to support history with low-overhead, bulk processing engines can not efficiently store the worker-results because of processing coarse tasks. 



 In principle, with massive system engineering efforts, machine learning practitioners can implement one-off asynchronous optimization methods with re-engineering  systems and interfaces. However, this comes at the cost of pushing system challenges such as scheduling, bookkeeping, and fault tolerance to the application developer. For example, Spark can support history if previous worker-results are stored to disk and checkpointed; this will create large storage overheads. Supporting CCMs is more challenging as the entire Spark engine has to change to support asynchronous execution.   An expert MPI programmer can use asynchronous primitives to implement SSP \cite{graham2005open}. However, this leads to increased program complexity and the complexity will increase if customized consistency models were to be implemented. Noteworthy, MPI does not have a robust support for fault tolerance and thus is typically not used for cloud computing. 


This work presents \framework{}, a bulk processing cloud-computing framework, built on top of  Spark,  that supports the implementation of distributed optimization methods with asynchrony and history. \framework{} implements an asynchronous execution to Spark's engine and enables the workers and/or the master to \textit{bookkeep} (log)  system-specific and application parameters. The asynchronous execution paradigm and the bookkeeping structures work together to construct a dynamic dependence graph for the implementation of custom consistency models and to recover history with a partial broadcast of model parameters.
Major contributions of this paper are:

    \begin{itemize}
        \item A novel framework for machine learning practitioners to implement and dispatch asynchronous machine learning applications  with custom consistency models on cloud and distributed platforms. \framework{} introduces three modules to cloud engines, \textit{ASYNCcoodinator}, \textit{ASYNCbroadcaster}, and \textit{ASYNCscheduler} to enable the asynchronous gather, broadcast, and schedule of tasks and results.
         
        \item A efficient history recovery strategy implemented with the \textit{ASYNCbroadcaster} and bookkeeping \textit{attributes},  to facilitate the implementation of variance reduced optimization methods that operate on historical gradients. 
        
        \item A robust programming model with extensions to the Spark API that enables the implementation of asynchrony and history while preserving the in-memory and fault tolerant features of Spark.  
        
        \item A demonstration of ease-of-implementation in \framework{} with the implementation and performance analysis of the  stochastic gradient descent (SGD) \cite{bottou2012stochastic}  algorithm and its asynchronous variant using a CCM. Also, the implementation of the history-based optimization method SAGA \cite{defazio2014saga} and its asynchronous variant in \framework{}. Our results demonstrate that asynchronous SAGA (ASAGA) \cite{leblond2016asaga} and asynchronous SGD (ASGD) outperform their synchronous variants up to \mg{4} times  on a distributed system with stragglers.
    \end{itemize}

\mg{\section{Preliminaries}}


Distributed machine learning often results in solving an optimization problem in which an objective function is optimized by iteratively updating the \textit{model parameters} until convergence. Distributed implementation of optimization methods includes workers that are assigned tasks to process parts of the training data, and one or more servers, i.e. masters, that store and update the model parameters. \mg{Distributed} machine learning models often result in 
the following structure:

\begin{equation}\label{opt-pbm}
    \min_{w \in \mathbb{R}^d} F(w) = \mg{\frac{1}{m}}\sum_{i=1}^\mg{m} {f^{(i)}(w)}
\end{equation}
where $w$ is the model parameter to be learned, $m$ is the number of \mg{workers}, and $f^{(i)}(w)$ is the \mg{local} loss function computed by \mg{worker } \textit{i} based on its \mg{assigned training data}. Each worker has access to $n_i$ data points, where the local cost has the form

\begin{equation}\label{loss}
  {f^{(i)}(w) := \sum_{j=1}^{n_i} \bar{f}^{(i)}_{j}(w)}
 \end{equation}

\noindent
for some loss functions $\bar{f}^{(i)}_{j}:\mathbb{R}^d \to \mathbb{R}$. \mg{For example, in supervised learning, given an input-output pair $\big(x_{ij}, y_{ij}\big)$, the loss function can be $\bar{f}_j^i(w) = \ell( \langle w,\phi(x_{ij})\rangle, y_{ij})$ where $\phi$ is a fixed function of choice and  $\ell(\cdot,\cdot)$ is a convex loss function that measures the loss if $y_{ij}$ is predicted from $x_{ij}$ based on the model parameter $w$. This setting covers empirical risk minimization problems in machine learning that include linear and non-linear regression, and other classification problems such as logistic regression \cite{reddi2015variance}. In particular, if $\phi(x)=x$ and the $\ell(\cdot,\cdot)$ function is the square of the Euclidean distance function, we obtain the familiar least squares problem
\begin{equation}
    \bar{f}_j^i(w) = \| x_{ij}^T w - y_{ij}\|^2
\end{equation}
where 
\begin{equation}\label{def-linear-reg}
  {f^{(i)}(w) := \sum_{j=1}^{n_i} \bar{f}^{(i)}_{j}(w)} = \| A_i w - b_i\|^2 
 \end{equation}
with $b_i = \{y_{ij}\}_{j=1}^{n_i}$ is a column vector of length $n_i$ and $A_i\in \mathbb{R}^{n_i \times d}$ is called the \emph{data matrix} as its $j$-th row is given by the input $x_{ij}^T$}. 
%


In the following we use the gradient descent (GD) algorithm as an example to introduce stochastic optimization and other terminology used throughout the paper such as \textit{mini-batch} size.
The introduced terms are used in all optimization problems and are widely used in the machine learning literature.  
GD iteratively computes the gradient of the loss function $\nabla F(w_k) =  \frac{1}{m}\sum_{i=1}^m \nabla f^{(i)}(w_k)$  to update the model parameters at iteration $k$.   To implement gradient descent on a distributed system, each worker $i$ computes its \emph{local gradient} $\nabla f^{(i)}(w^k)$; the local gradients are aggregated by the master when ready. The full pass over the data at every iteration of the algorithm with synchronous updates leads to large overheads. Distributed stochastic gradient descent (SGD) methods and their variants \cite{chen2016revisiting} are on the other hand scalable and popular methods for solving (\ref{opt-pbm}). Distributed SGD replaces the local gradient $\nabla f^{(i)}(w_k)$ with an unbiased stochastic estimate $\tilde\nabla f^{(i)}(w_k)$ of it, computed from a subset of local data points: 
\begin{align}
\label{eqn:stoch_grad}
\nabla \tilde{f}^{(i)} (w_k) :=  \frac{1}{b_i} \sum\limits_{s \in S_{i,k}}  \nabla \bar {f}_s^{(i)} (w_k),
\end{align} 
where $S_{i,k} \subset \{1,\dots,n_i\}$ is a random subset that is sampled with or without replacement at iteration $k$, and $b_i := |S_{i,k}|$ is the number of elements in $S_{i,k}$ \cite{bottou2012stochastic}, also called the 
mini-batch size. To obtain desirable accuracy and performance, implementations of stochastic optimization methods  require tuning algorithm parameters. For example, the step size and the mini-batch sizes are parameters to tune in SGD \cite{bottou2012stochastic}.
%

\begin{figure*}
\centering
  \includegraphics[width=0.6163\linewidth]{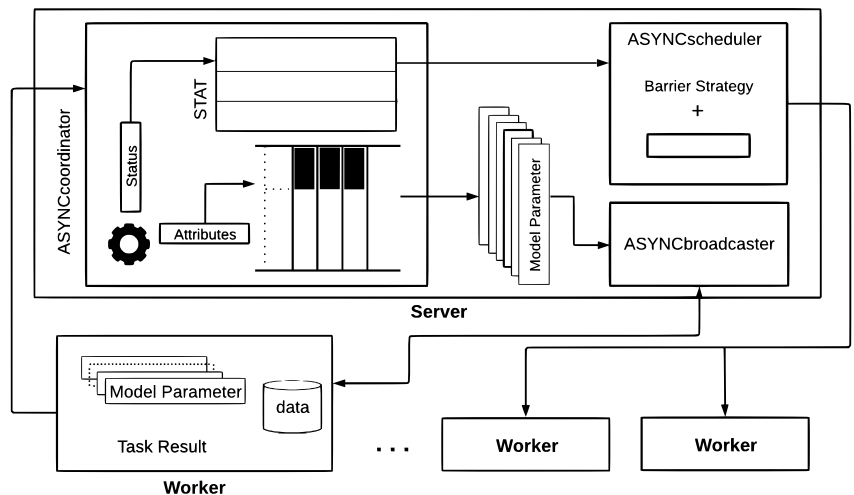}
  \caption{An overview of the ASYNC framework.}
  \label{fig:framework_fig}
\end{figure*}

\vspace{0.05in}
\mg{\section{Motivation for Asynchrony and History}}

Asynchrony is implemented to improve the converge rate and time-to-solution of optimization methods on cluster-computer platforms with slow machines (stragglers).   
In distributed optimization, workers compute local gradients of the objective function and then  communicate the computed gradients to the server. To proceed to the next iteration of the algorithm, the server updates the shared model parameters with the received gradients, broadcasts the most recent model parameter, and schedules  new tasks.  
 In asynchronous optimization, the server can proceed with the update and broadcast of the model parameters without having to wait for all worker tasks to complete. This asynchrony allows the algorithm to make progress in the presence of stragglers which is known as an increase in hardware efficiency \cite{cipar2013solving}. However, this progress in computation comes at a cost, the asynchrony inevitably adds \textit{staleness} to the system wherein some of the workers compute gradients using model parameters that may be several gradient steps behind the most updated set of model parameters which can lead to poor convergence. This is also referred to as a worsening in statistical efficiency \cite{chen2016revisiting}. 
 
 Asynchronous optimization methods are formulated and implemented with properties that balance statistical efficiency and hardware efficiency to maximize the performance of the optimization methods on distributed systems. Consistency models, i.e. barrier control strategies, are used to design asynchronous optimization methods that enable this balance.
%
Barriers in asynchronous algorithms determine if a worker should proceed to the next iteration or if it should wait until a specific number of workers have communicated their results to the server.
%
The most well-known barrier control strategy is the  Stale Synchronous Parallel (SSP)  in which workers synchronize when staleness (determined by the number of stragglers) exceeds a threshold. 
\framework{} supports SSP and also facilitates the implementation of custom consistency models that apply barriers based on parameters such as worker-task-completion time and scheduling delays. 

History augments the noise from stochastic gradients to improve the convergence rate of the optimization method. Distributed optimization methods used in machine learning applications are typically stochastic  \cite{bottou2012stochastic}. Stochastic optimization methods use a noisy gradient computed from random data samples instead of the true gradient which can lead to poor convergence. Variance reduction techniques, used in both synchronous and asynchronous optimization, augment the noisy gradient to reduce this variance. A class of variance-reduced asynchronous algorithms that have led to significant improvements over traditional methods memorize the gradients computed in previous iterations, i.e. historical gradients \cite{defazio2014saga}. Historical gradients can not be implemented in cluster-computing engines such as Spark primarily because Spark can only broadcast the entire history of the model parameters which can be very large and can lead to significant overheads. 


\begin{table*}
\centering
\caption{Transformations, actions, and  methods  in \framework{}. AC is the ASYNCcontext and Seq[T] is a sequence of elements of type T.}
\label{tab:functions}
\begin{tabular}{|l|l|l|} 
\hline
Actions         & \begin{tabular}[c]{@{}l@{}}ASYNCreduce(f:(T,T) $\Rightarrow$ T, AC)\\\\ASYNCaggregate(zeroValue: U) \\(seqOp: (U, T) $\Rightarrow$ U, combOp: (U, U) $\Rightarrow$ U), AC)\end{tabular} & \begin{tabular}[c]{@{}l@{}}Reduces the elements of the RDD using the specified associative\\ binary operator.\\Aggregates the elements of the partition using the combine  \\ functions and a neutral "zero" value.\end{tabular}\\ 
\hline
Transformations & ASYNCbarrier(f:T $\Rightarrow$ Bool, Seq[T])& Returns a RDD containing elements that satisfy a predicate \textit{f}.\\ 
\hline
Methods         & \begin{tabular}[c]{@{}l@{}}ASYNCcollect()\\ASYNCcollectAll()\\ASYNCbroadcast(T)\\AC.STAT\\AC.hasNext() \end{tabular}                              & \begin{tabular}[c]{@{}l@{}}Returns a task result.\\Returns a task result and its attributes.\\
Creates a dynamic broadcast variable.\\Returns the current status of all workers.\\Returns true if a task result exists.\end{tabular}  \\
\hline
\end{tabular}
\end{table*}

\vspace{0.05in}

\section{\framework{}: A Cloud Computing Framework with Asynchrony and History }

\framework{} is a framework, built on top of Spark \cite{zaharia2012resilient}, for the implementation and execution of asynchrony and history in optimization algorithms while retaining the map-reduce model, scalability, and fault tolerance of state-of-the-art cluster computing engines. 
%
%
%
%
\autoref{fig:framework_fig} demonstrates an overview of the \framework{} engine.  The three main modules in \framework{} are the \textit{ASYNCcoordinator}, \textit{ASYNCbroadcaster}, \textit{ASYNCscheduler}. \framework{} also collects and stores \textit{bookkeeping structures}. These structures are communicated between the workers and the master and are either system-specific, i.e. \textit{status}, or are related to the application, i.e. \textit{attributes}. This section elaborates how the internal elements of \framework{}  work together to facilitate the implementation of asynchrony and history. 

 \textit{Bookkeeping structures in \framework{}.} Bookkeeping structures  are used by the main modules of \framework{}  to enable the implementation of asynchrony and history.  
 These structures are collected by \framework{} at runtime and are stored on the master. 
 With the help of the ASYNCcoordinator, each worker communicates to the master, application-specific attributes such as task results and the mini-batch size. Workers' recent status such as worker staleness, average-task-completion time, and availability\footnote{A worker is available if it is not executing a task and unavailable otherwise} are also logged and stored in a table called STAT with the help of the ASYNCcoordinator.  

\color{black}

\textit{Implementing asynchrony with the ASYNCcoordinator, ASYNCscheduler, and the status structures.}
To implement asynchrony, \framework{} implements a dynamic task graph computation model which uses the consistency model to dynamically determine executing tasks and their assigned workers. The execution of tasks on workers is automatically triggered by the system using a computation graph. Task and data objects are the nodes in this graph and the edges are the dependency amongst nodes/tasks.   The computation graph in classic consistency models such as SSP does not change at runtime because the models do not rely on runtime information such as the system state. However, many CCMs take  information from the current state of the system as input and couple this information with the barrier control strategy to dynamically build the computation graph. 
To implement CCMs, the ASYNCcoordinator periodically communicates with the workers to update system-specific parameters in the STAT table. The ASYNCscheduler uses the parameters in STAT and a user-defined barrier control strategy to update the computation graph. The computation graph is then executed to apply the desired constancy model.  

\textit{Implementing history with the ASYNCbroadcaster and the attributes.}
In each iteration of an optimization method that uses history, the computed gradients from previous iterations are used together with the current model parameters to update the model parameters. 
Implementing history in a coarse-grained computation engine via  explicitly storing bulky worker-results, i.e. previous gradients,   leads to significant storage overheads. A fault tolerant execution will also have overheads in this approach as large gradients have to be periodically check-pointed or recomputed explicitly using a lineage. 

\framework{} does not explicitly store, communicate, or compute past gradients. Instead we use the approach from \cite{reddi2015variance} in which  the history of past gradients is recovered, when needed, using previous model parameters. 
Recovering history has low storage and computation overheads in coarse-grained computation models. By recovering history, workers in \framework{} do not need to store any previously computed gradients and only the previous model parameters are stored on the master. The cost of storing the model parameters has an inverse relation to the batch size \cite{bottou2012stochastic} and thus reduces as the granularity of tasks increase, e.g. larger batch sizes.  Also, to recover a past gradient, a worker only needs to subtract its recent model parameter from the previous model parameter that is broadcasted to it from the master; the approach in \cite{reddi2015variance} is then used to update the master-side model parameters based on the history.  
The ASYNCbroadcaster in \framework{} is responsible for the asynchronous broadcast of model parameters between the master and individual workers. 
Attributes such as the mini-batch size, required by the master to apply history to its model parameters, are also broadcast using the ASYNCbroadcaster.

\color{black}

\section{Programming with \framework{}}
To use \framework{}, developers are provided an additional set of \framework{}-specific functions, on top of what Spark provides, to access the bookkeeping structures and to implement asynchrony and history. The programming model in \framework{} is close to that of Spark. It operates on resilient distributed datasets (RDD) to preserve the fault tolerant and in-memory execution of Spark. The \framework-specific functions also either transform  the RDDs, known as \textit{transformations} in Spark, or conduct lazy actions.   In this section, \framework{}'s programming model and API is first discussed. 
We then show the implementation of  SGD and its asynchronous variant which uses a CCM. A well-known history-based optimization method called SAGA \cite{defazio2014saga} and its asynchronous variant with a CCM is also implemented. Finally, we discuss the implementation of other consistency models in \framework{}.



\subsection{The \framework{} programming model}
Asynchronous Context (AC) is the entry point to \framework{} and should be created only once in the beginning of the application. The {\framework{}scheduler}, the {\framework{}broadcaster},  and the {\framework{}coordinator} communicate via the AC and with this communication create barrier controls, broadcast variables, and store  workers' task results and status. AC  maintains the bookkeeping structures  and \framework{}-specific functions, including  actions and transformations that operate on RDDs.
Workers use \framework{} functions to interact with AC and to store their results and attributes in the bookkeeping structures. The server queries AC to update the model parameters or to access workers' status.   \autoref{tab:functions} lists the main functions  available in \framework{}. We show the signature of each operation by demonstrating the type parameters in square brackets.


\textit{Collective operations in \framework{}}. \textit{\framework{}reduce} is an action that aggregates the elements of the RDD on the worker  and returns the result to the server. {\framework{}reduce} differs from Spark's \textit{reduce} in two ways. First, Spark aggregates data across each partition and then combines the partial results together to produce a final value. However,  {\framework{}reduce} executes only on the worker and for each partition. Secondly, \textit{reduce} returns only when all partial results are combined on the server, but {\framework{}reduce} returns immediately. Task results on the server are accessed using the \textit{\framework{}collect} and \textit{\framework{}collectAll} methods. \framework{}collect returns  task results in FIFO (first-in-first-out) order and also returns the worker  attributes.  The workers' status can also be accessed with \textit{\framework{}.STAT}.

\begin{algorithm}
\begin{small}
\SetKwInOut{Input}{Input}
\SetKwInOut{Output}{Output}
\Input{points, numIterations, learning rate $\alpha_i$, sampling rate \textit{b}}
\Output{ model parameter $w$}
\For{ i = 1 to numIterations }{
    w\_br = sc.broadcast(w)\\
    gradient = points.sample(b).map(p $\Rightarrow$  $\nabla f_p(w\_br.value)$). reduce(\_+\_)\\
    w -= $\alpha_i$ $\ast$ gradient \\
}
return $w$ 
\end{small}
\caption{ The SGD Algorithm}
\label{alg:sgd}
\end{algorithm}
\vspace{-0.1in}
\begin{algorithm}
\begin{small}
\SetKwInOut{Input}{Input}
\SetKwInOut{Output}{Output}
\Input{points, numIterations, learning rate $\alpha_i$,  sampling rate \textit{b}}
\Output{model parameter $w$}
\color{blue}  AC =  new ASYNCcontext \color{black}\\
\For{ i = 1 to numIterations }{
    w\_br = sc.broadcast(w)\\
    points.\color{blue} ASYNCbarrier(f, AC.STAT)\color{black}.sample(b).map(p $\Rightarrow$  $\nabla f_p(w\_br.value)$)
     \color{blue}.ASYNCreduce(\_+\_, AC) \color{black}\\
    \While{ \color{blue}\text{AC.hasNext()} \color{blue}}{
        gradient\color{blue} = AC.ASYNCcollect()\color{black}\\
        w -= $\alpha_i$ $\ast$ gradient \\
    }
}
return $w$ 
\end{small}
\caption{The ASGD Algorithm}
\label{alg:asgd}
\end{algorithm}
\vspace{-0.1in}

\begin{algorithm}
\begin{small}
\SetKwInOut{Input}{Input}
\SetKwInOut{Output}{Output}
\Input{points, numIterations, learning rate $\alpha$,  sampling rate \textit{b}, number of points \textit{n}}
\Output{model parameter $w$}
averageHistory = 0\\
store $w$ in table\\
\For{ i = 1 to numIterations }{
    w\_br =sc.broadcast(w)\\
    (gradient, history)= points.sample(b).map((index,p) $\Rightarrow$  $\nabla {f_p(w\_br.value)}$,  $\nabla {f_p(\color{red}table[index]\color{black})}$).reduce(\_+\_)\\
    averageHistory += (gradient - history)$\ast$ b$\ast$n\\
    w -= $\alpha$ $\ast$ (gradient - history + averageHistory ) \\
    update table\\
}
return $w$ 
\end{small}
\caption{The SAGA Algorithm}
\label{alg:saga}
\end{algorithm}

\begin{algorithm}
\begin{small}
\SetKwInOut{Input}{Input}
\SetKwInOut{Output}{Output}
\Input{points, numIterations, learning rate $\alpha$,  sampling rate \textit{b}, \#points \textit{n}, \#partitions P}
\Output{model parameter $w$}
\color{blue}  AC =  new ASYNCcontext \color{black}\\
averageHistory = 0\\
\For{ i = 1 to numIterations }{
    w\_br = \color{blue}AC.ASYNCbroadcast(w)\color{black}\\
     points.\color{blue}ASYNCbarrier(f, AC.STAT) \color{black}.sample(b).map((index,p) $\Rightarrow$  $\nabla {f_p(w\_br.value )}$, $\nabla {f_p( w\_br.{\color{blue}value(index) \color{black}})}$). \color{blue}ASYNCreduce(\_+\_, AC)\color{black}\\
    \While{\color{blue}AC.hasNext()\color{black}}{
        (gradient,history)\color{blue} = AC.ASYNCcollect()\color{black}\\
        averageHistory += (gradient - history)$\ast$ b$\ast$n/P\\
        w -= $\alpha$ $\ast$ (gradient - history + averageHistory ) \\
    }
}
return $w$ 
\end{small}
\caption{The ASAGA Algorithm}
\label{alg:asaga}
\end{algorithm}

\textit{Barrier and broadcast in \framework{}.} \textit{\framework{}barrier} is a \textit{transformation}, i.e. a deterministic operation which creates a new RDD based on the workers' status.  {\framework{}barrier} takes  the recent status of workers.\textit{STAT} and decides which workers to assign new tasks to, based on a user-defined function. For example, for a fully asynchronous barrier model  the following function is declared: $f: STAT.foreach(true)$. 
In Spark, broadcast parameters are ``broadcast variable'' objects that wrap around the to-be-broadcast value. \textit{\framework{}Broadcast} also uses broadcast variables and similar to Spark the method \textit{value} can be used to access the broadcast value.
However, ASYNCbroadcast differs from the broadcast implementation in Spark since it has access to an \textit{index}. The index is used internally by ASYNCbroadcast to get the ID of the previously broadcast variables for the specified index. ASYNCbroadcast eliminates the need  to broadcast values when accessing the history of broadcast values.

%


\subsection{Case studies}
The robust programming model in \framework{} provides control of low-level features in both the algorithm and the execution platform to facilitate the implementation of asynchrony and history in optimization methods.  The following demonstrates the implementation of  well-known asynchronous optimization methods ASGD and ASAGA in \framework{} as examples.

\textit{ASGD with \framework{}}. An implementation of mini-batch stochastic gradient descent (SGD) using the map-reduce model in Spark is shown in Algorithm \ref{alg:sgd}. The map phase applies the gradient function on the input data independently on workers. The reduce phase has to wait for all the map tasks to complete. Afterwards, the server aggregates the task results and updates the model parameter \textit{w}. The asynchronous implementation of SGD in \framework{} is shown in Algorithm \ref{alg:asgd}. With only a few extra lines from the \framework{} API, colored in blue, the synchronous implementation of SGD in Spark is transformed to ASGD. An {\framework{}context} is created in line 1 and  is used in line 4 to create a barrier using the user-defined CCM indicated by $f$ and based on the current workers' status, {AC.STAT}.   The partial results from each partition are  then obtained and stored in AC in line 4. Finally, these partial results are accessed in line 6 and are used to update the model parameter in line 7.

 
%


\textit{ASAGA with \framework{}}. The SAGA implementation in Spark is shown in Algorithm \ref{alg:saga}. This implementation is inefficient and not practical for large datasets as it needs to synchronously broadcast a table of all stored model parameters to each worker, colored in red in Algorithm \ref{alg:saga} line 5. The size of this increases after each iteration and thus  broadcasting it leads to large communication overheads. As a result of the overhead, machine learning libraries that are build on top of Spark such as Mllib \cite{meng2016mllib} do not provide implementations of optimization methods such as SAGA that requires the history of gradients. \framework{} resolves the overhead with \framework{}broadcast.  The implementation of ASAGA  is shown in Algorithm \ref{alg:asaga}. {\framework{}broadcast} is used to define a dynamic broadcast in line 4. Then, the broadcast variable is used to compute the historical gradients in line 5. In order to access the last model parameters for sample \textit{index}, the method \textit{value} is called in line 5. As shown in Algorithm \ref{alg:asaga}, there is no need to broadcast a table of parameters which allows for  efficient implementation of both SAGA and ASAGA in \framework{}.




\textit{CCMs in \framework{}}. To enable the implementation of custom consistancy models, \framework{} provides the interface to implement user-defined functions that  selectively choose from available workers based on their status. \autoref{list:barrier} demonstrates the implementation of two CCMs in  \framework{}, CCM1 and CCM2, as well as the SSP model. CCM1 is the throttled-release \cite{albrecht2006loose} barrier strategy which submits tasks to available workers only when the  number of available workers is at least $k$. CCM2 implements a fully asynchronous barrier that allows workers to progress as soon as their current task finishes. 
\begin{lstlisting}[label={list:barrier},caption={Pseudo-code for implementing CCMs in ASYNC.},captionpos=b]
f: STAT.foreach(Avaialble_Workers >= k) % @CCM1@ 
f: STAT.foreach(true) %  @CCM2@ 
f: STAT.foreach(MAX_Staleness < s) % The @SSP@ barrier control with a staleness threshold 's'
points.ASYNCbarrier(f, AC.STAT) % Apply the barrier 
\end{lstlisting}

\section{Results}

We evaluate the performance of \framework{} by implementing two asynchronous optimization methods, namely ASGD and ASAGA,  and their synchronous variants to solve {least squares problems}. We implement the
throttled-release CCM for the both asynchronous methods and use history in ASAGA and SAGA.   The performance of ASGD and ASAGA are compared to their synchronous implementations in Spark. To the best of our knowledge, no library or implementation of asynchronous optimization methods exists on Spark. However, to demonstrate that the synchronous implementations of the algorithms using \framework{} are well-optimized, we first compare the performance of the synchronous variants of the tested optimization methods in \framework{} with the state-of-the-art machine learning library, Mllib \cite{meng2016mllib}. Mllib is a library that provides implementations of a number of synchronous optimization methods. 
In \autoref{sec:straggler} we evaluate the performance of ASGD and ASAGA in \framework{} in the presence of stragglers.  

\begin{table}
  \centering
 \begin{tabular}{ |p{1.8cm}|p{1.8cm}|p{2.1cm}|p{1.3cm}|}
 \hline
 Dataset & Row numbers & Column numbers & Size\\
 \hline
rcv1\_full.binary &  697,641 & 47,236 & 851.2MB\\
mnist8m &  8,100,000 & 784 & 19GB\\
epsilon &  400,000 & 2000 &  12.16GB\\
 \hline
\end{tabular}
  \caption{Datasets for the experimental study.}
  \label{tab:datasets}
\end{table}

\begin{figure}
  \includegraphics[width=0.94\linewidth]{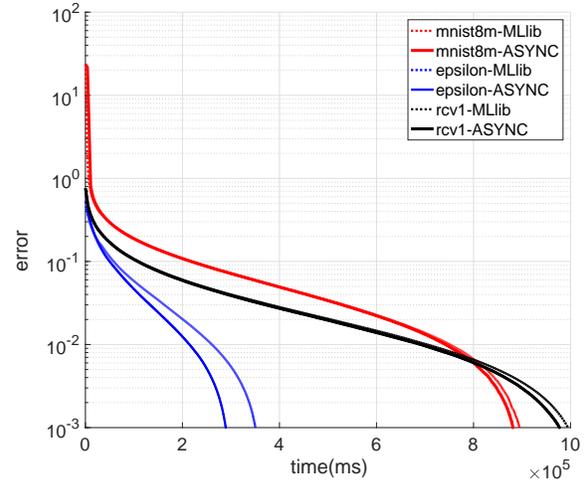}
  \caption{The performance of SGD implemented in \framework{} versus Mllib.}
  \label{fig:asgd-compareWithMllib}
\end{figure}

\begin{figure*}
    \begin{subfigure}[b]{0.34\textwidth}
        \includegraphics[width=\textwidth]{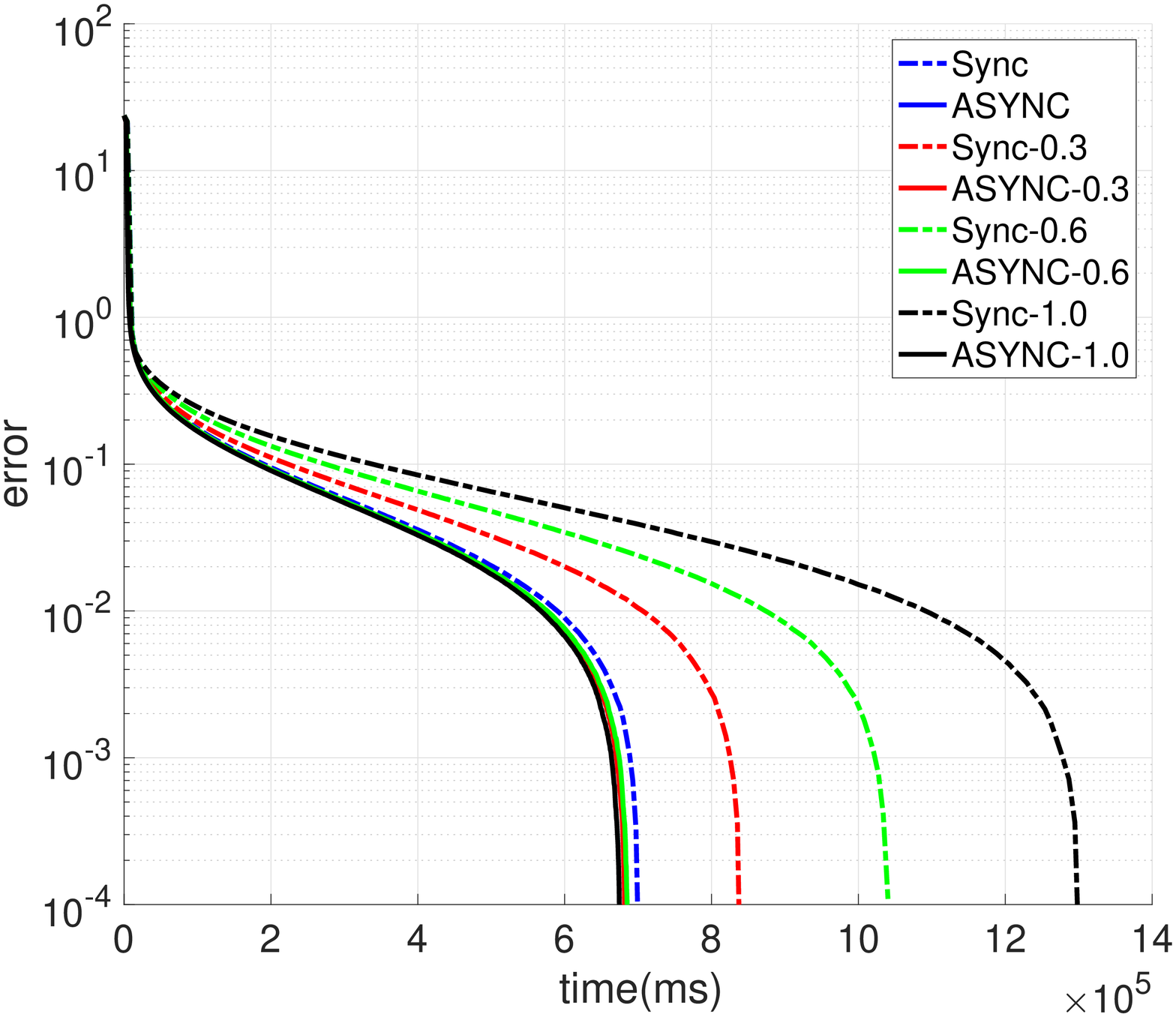}
        \caption{mnist8m }
        \label{fig:asgd-mnist8m}
    \end{subfigure}
    ~ 
    \begin{subfigure}[b]{0.34\textwidth}
        \includegraphics[width=\textwidth]{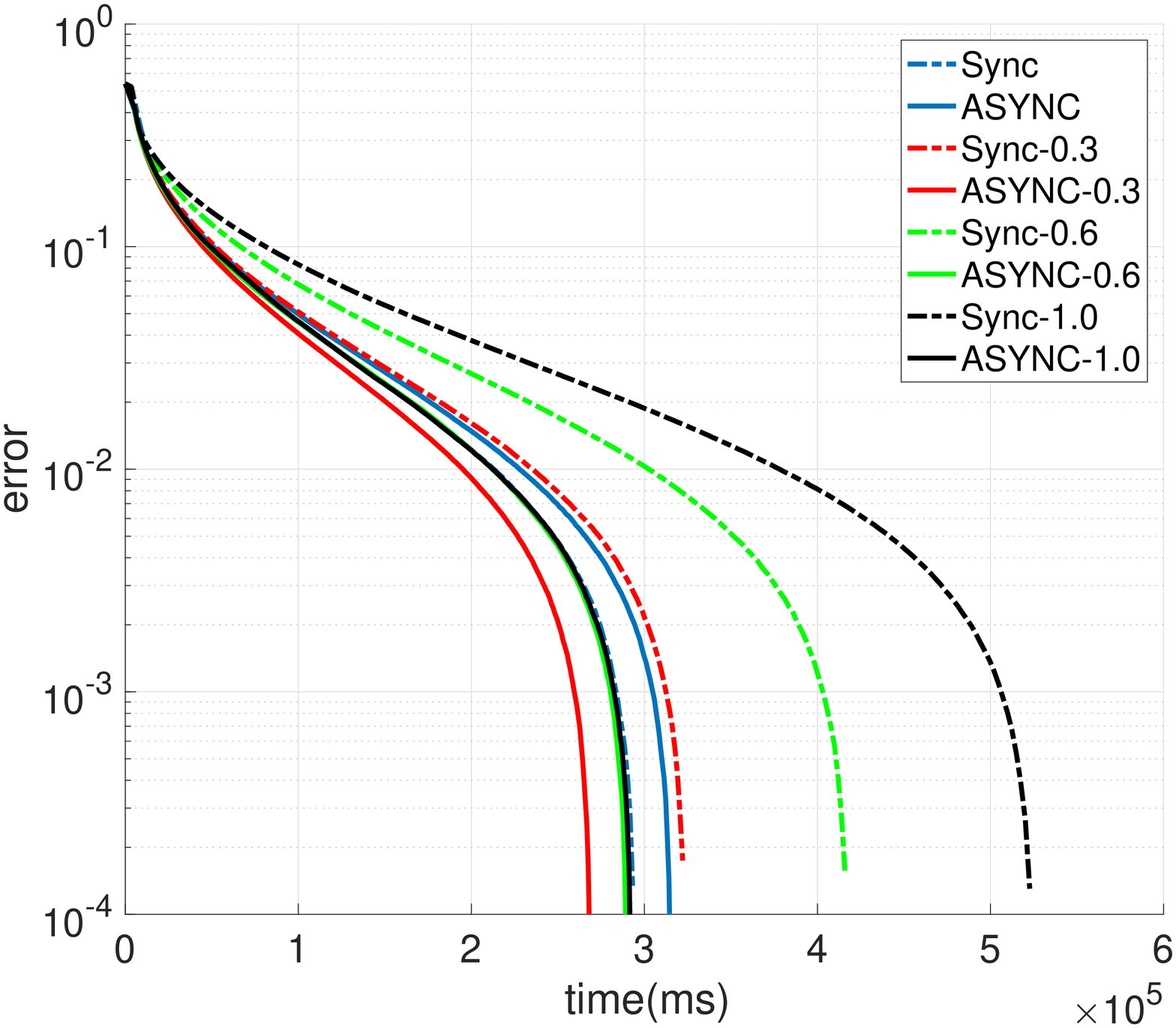}
        \caption{epsilon}
        \label{fig:asgd-epsilon}
    \end{subfigure}
    ~ 
    \begin{subfigure}[b]{0.34\textwidth}
        \includegraphics[width=\textwidth]{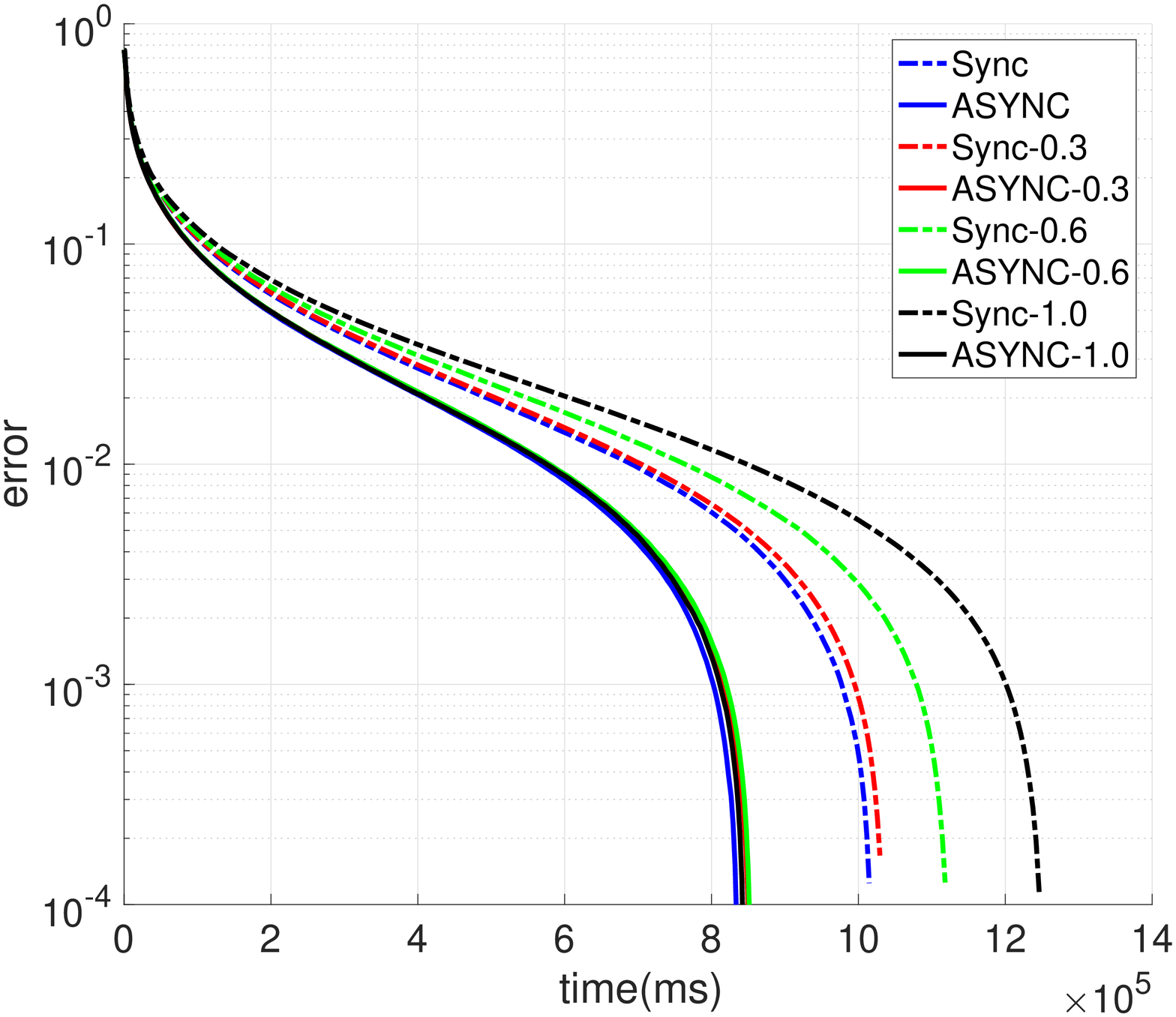}
        \caption{rcv1\_full.binary}
        \label{fig:asgd-rcv}
    \end{subfigure}
    \caption{The performance of ASGD and SGD in \framework{} with 8 workers for different delay intensities of 0\%, 30\%, 60\% and 100\% which are shown with ASYNC/SYNC, ASYNC/SYNC-0.3, ASYNC/SYNC-0.6 and ASYNC/SYNC-1.0 respectively.}\label{fig:asgd}
\end{figure*}

\subsection{Experimental setup}
\mg{We consider the distributed least squares problem defined in \eqref{def-linear-reg}}. Our experiments use the datasets listed in \autoref{tab:datasets} \mg{from the LIBSVM library \cite{chang2011libsvm}}, all of which vary in size and sparsity. 
\mg{The first dataset rcv1\_full.binary is about documents in the Reuters Corpus Volume I (RCV1) archive, which are newswire stories. The second dataset mnist8m contains handwritten digits commonly used for training various image processing systems, and the third dataset epsilon is the Pascal Challenge 2008 that predicts the presence/absence of an object in an image.}
\mg{For the experiments, we use} \framework{}, Scala 2.11, Mllib \cite{meng2016mllib}, and Spark 2.3.2. Breeze 0.13.2 and netlib 1.1.2 are used for the (sparse/dense) BLAS operations in \framework{}. \mg{ XSEDE Comet CPUs \cite{towns2014xsede} are used to assemble the cluster.

To demonstrate the performance of  asynchronous algorithms and their robustness to the heterogeneity in cloud environments, we evaluate the implemented methods in the presence of stragglers.   
Two different straggler behaviours are used: \textit{(i)} \textit{Controlled Delay Straggler (CDS)} experiments in which a single worker is delayed with different intensities; \textit{(ii)} the \textit{Production Cluster Stragglers (PCS)} experiments in which straggler patterns from real production clusters are used.   The CDS experiments  are ran with all three datasets on a cluster composed of a server and 8 workers. The PCS experiments require a larger cluster and thus are conducted on a cluster of 32 workers with one server using the two larger datasets (mnist8m and epsilon). In all configurations a worker runs an executor with 2 cores.}
The number of data partitions is 32 for all datasets and in the implemented algorithms. \mg{The experiments are repeated three times; the average reported.}

\textit{Parameter tuning}: A sampling rate of \textit{b = 10\%} is selected  for the mini-batching SGD for \textit{mnist8m} and \textit{epsilon} and \textit{b = 5\%} is used for \textit{rcv1\_full.binary}. SAGA and ASAGA use \textit{b = 10\%} for \textit{epsilon}, \textit{b = 2\%} for \textit{rcv1\_full.binary}, and use \textit{b =1\%} for \textit{mnist8m}.
\mg{For the PCS experiment, we use \textit{b = 1\%} for \textit{mnist8m} and \textit{epsilon}.}
%
%
We use the same step size as Mllib and tune it for SGD to converge faster. A fixed step size is used in SAGA which is also tuned for  faster convergence.  The step size is not tuned for the asynchronous algorithms. Instead, we use the following heuristic, the step size of ASGD and ASAGA is computed by dividing the initial step size of their synchronous variants by the number of workers \cite{recht2011hogwild}. \mg{We run the SGD algorithm in Mllib for 15000 iterations with sampling rate of 10\% and use its final objective value as the \textit{baseline} for the least squares problem.} 

\subsection{Comparison with Mllib}
We use \framework{} for implementations of both the synchronous and the asynchronous variants of the algorithms because \textit{(i)} \framework{}'s performance for synchronous methods is similar to that of  Mllib's; \textit{(ii)} asynchronous methods are not supported in Mllib; \textit{(iii)} synchronous methods that require history of gradients can not be implemented in Mllib because of discussed overheads.  To demonstrate that our implementations in \framework{} are optimized, we compare the performance of SGD in \framework{} and Mllib for solving the least squares problem \cite{svrg}.  Both implementations use the same initial step size. The  \textit{error} is defined as \textit{objective function value} minus the \textit{baseline}.  \autoref{fig:asgd-compareWithMllib} shows the error for three different datasets.  The figure demonstrates that SGD in \framework{} has a similar performance to that of Mllib's on 8 workers, the same pattern is observed on 32 workers. Therefore, for the rest of the experiments, we compare the asynchronous and synchronous implementations in \framework{}.



 

\begin{figure}
  \includegraphics[width=0.85\linewidth]{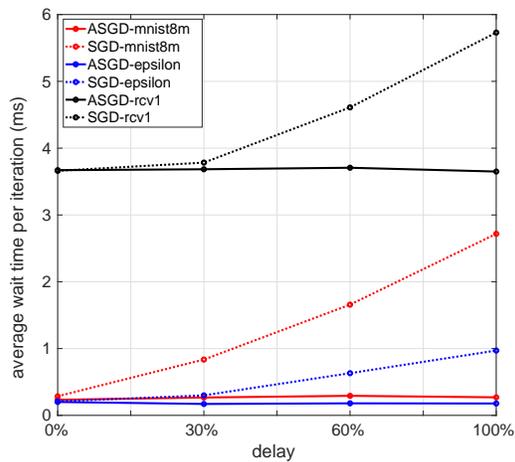}
  \caption{Average wait time per iteration with 8 workers for ASGD and SGD in ASYNC for different delay intensities.}
  \label{fig:asgd-wait_time}
\end{figure}

\begin{figure*}
    \begin{subfigure}[b]{0.343\textwidth}
        \includegraphics[width=\textwidth]{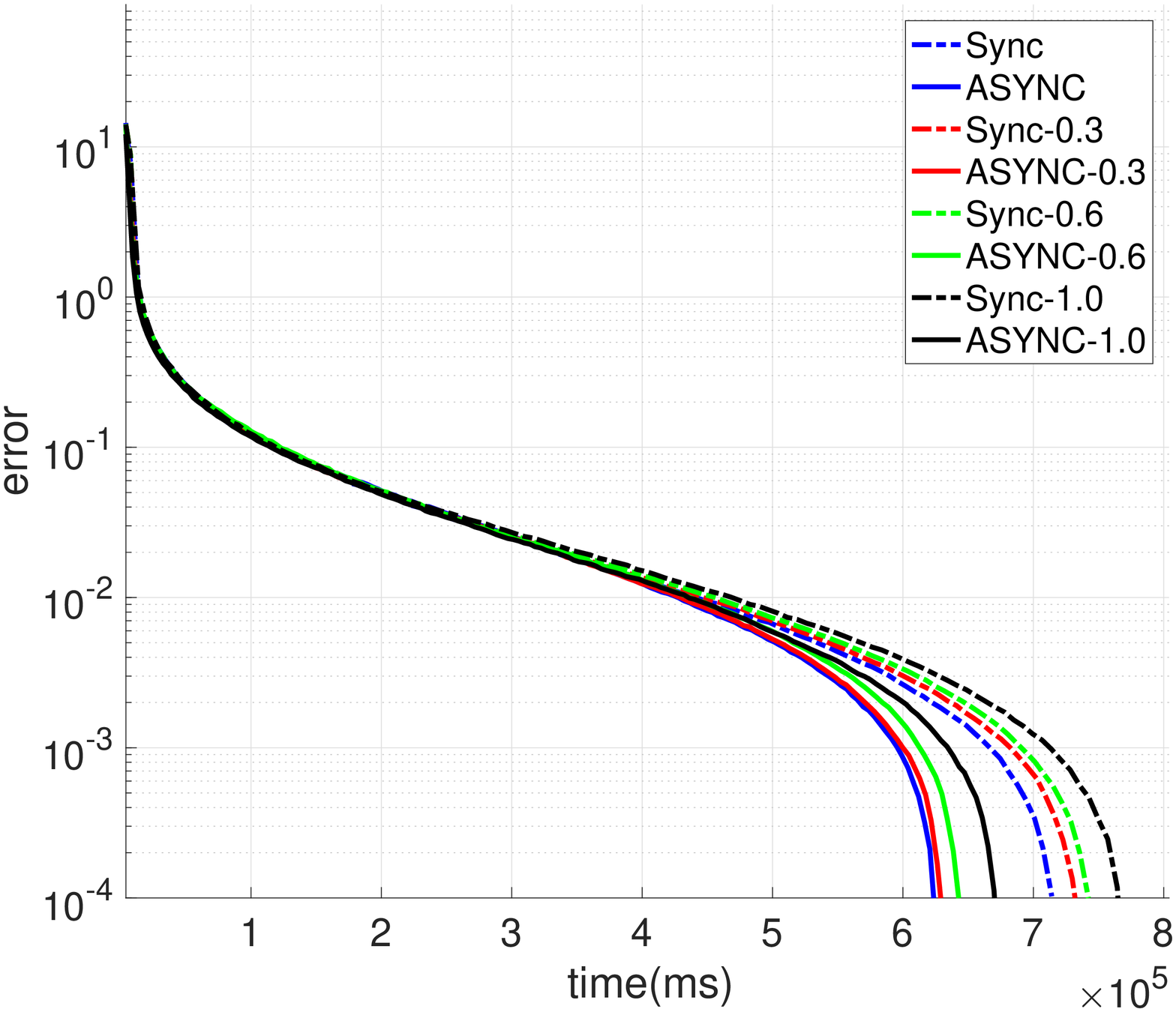}
        \caption{mnist8m }
        \label{fig:asaga-mnist8m}
    \end{subfigure}
    ~ 
    \begin{subfigure}[b]{0.343\textwidth}
        \includegraphics[width=\textwidth]{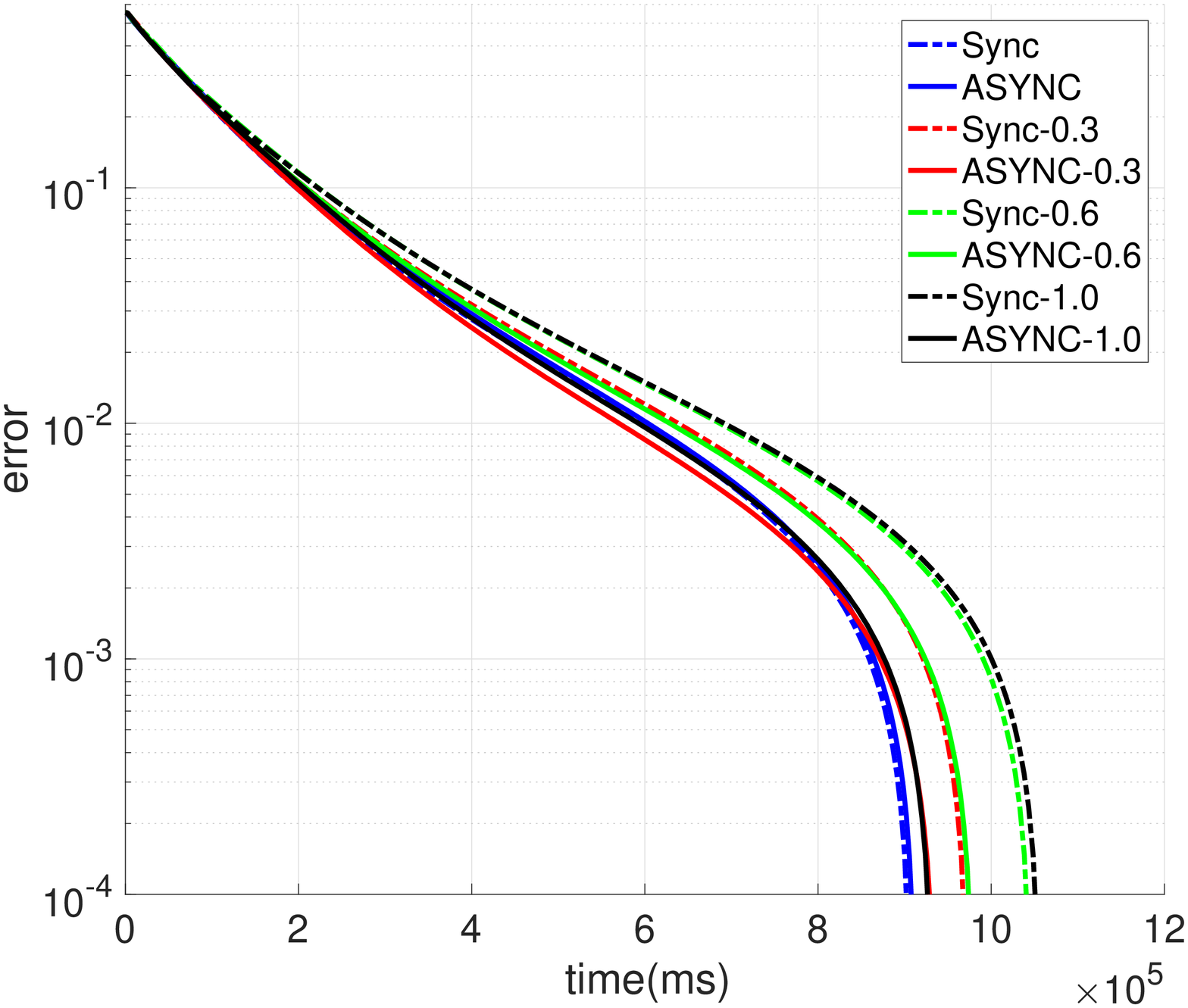}
        \caption{epsilon}
        \label{fig:asaga-epsilon}
    \end{subfigure}
    ~ 
    \begin{subfigure}[b]{0.343\textwidth}
        \includegraphics[width=\textwidth]{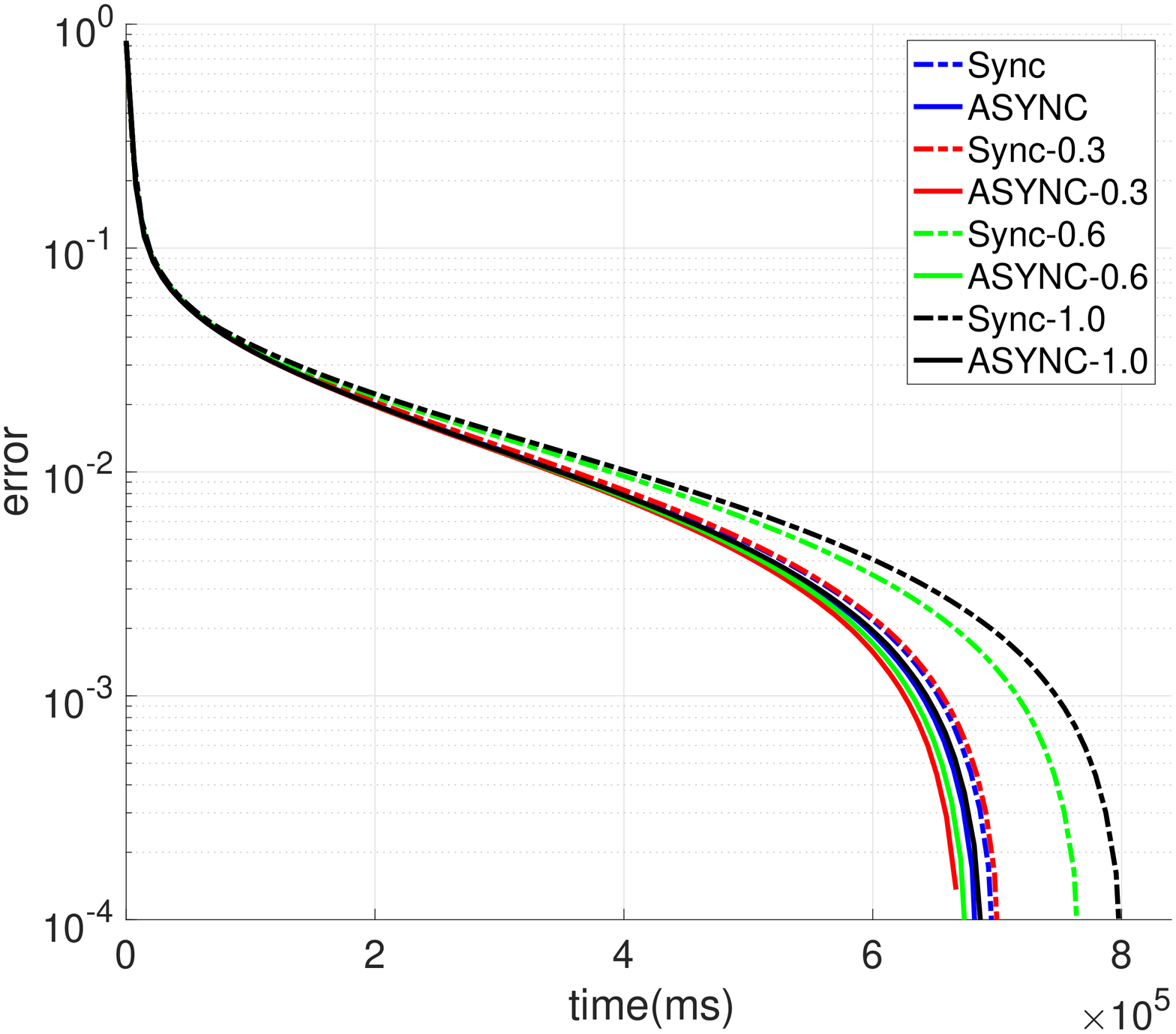}
        \caption{rcv1\_full.binary}
        \label{fig:asaga-rcv}
    \end{subfigure}
    \caption{The performance of ASAGA and SAGA in \framework{} for different delay intensities of 0\%, 30\%, 60\% and 100\% which are shown with ASYNC/SYNC, ASYNC/SYNC-0.3, ASYNC/SYNC-0.6 and ASYNC/SYNC-1.0 respectively.}\label{fig:asaga}
\end{figure*}

\subsection{Robustness to stragglers}
\label{sec:straggler}
\textit{Controlled Delay Straggler:} We demonstrate the effect of different delay intensities in a single worker on SGD, ASGD, SAGA, and ASAGA by simulating  a straggler with controlled delay \cite{karakus2017straggler,cipar2013solving}. From the 8 workers in the cluster,  a 
%
%
delay between 0\% to 100\% of the time of an iteration is added to one of the workers. The delay intensity, which we show with \textit{delay-value} \%, is the percentage by which a worker is delayed, e.g. a 100\% delay means the worker is executing jobs at half speed. The controlled delay is implemented with the sleep command. The first 100 iterations of both the synchronous and asynchronous  algorithms are used to measure the average iteration execution time.

\begin{table}[h]
\centering
\begin{tabular}{|l|l|l|l|l|}
\hline
        & SAGA    & ASAGA  & SGD    & ASGD   \\ \hline
mnist8m & 42.8367 ms & 9.8125 ms & 6.4433 ms & 3.5745 ms \\ \hline
epsilon & 6.9926  ms & 1.1721 ms & 5.3112 ms & 1.4165 ms\\ \hline
\end{tabular}
\caption{Average wait time per iteration on 32 workers. }
\label{tab:all-32}
\end{table}

The performance of SGD and ASGD for different delay intensities are shown in \autoref{fig:asgd} where for the same delay intensity  the asynchronous implementation always converges faster to the optimal solution compared to the synchronous variant of the algorithm. As the delay intensity increases, the straggler has a more negative effect on the runtime of SGD. However, ASGD converges to the optimal point with almost the same rate for different delay intensities. This is because the ASYNCscheduler continues to assign tasks to workers without having to wait for the straggler. When the task result from the straggling worker is ready, it independently updates  the model parameter. Thus, while ASGD in \framework{} requires more iterations to converge, its overall runtime is considerably faster than the synchronous method. With a delay intensity of \%100, a speedup of up to  2$\times$ is achieved with ASGD vs. SGD. 








 \autoref{fig:asgd-wait_time} shows the average wait time for each worker over all iterations for SGD and ASGD. The wait time is defined as the time from when a worker submits its task result to the server until it receives a new task. In the asynchronous algorithm, workers proceed without waiting for stragglers. Thus the average wait time does not change with changes in delay intensity. However, in the synchronous implementation worker  wait times increase with a slower straggler. For example, for the \textit{mnist8m} dataset in \autoref{fig:asgd-wait_time}, the average wait time for SGD increases significantly when the straggler is two times slower (delay = 100\%). Comparing  \autoref{fig:asgd} with \autoref{fig:asgd-wait_time} shows that the overall runtime of ASGD and SGD is directly related to their average wait time where an increase in the wait time negatively affects the algorithms convergence rate.


The slow worker pattern used for the ASGD experiments is also used for ASAGA.  \autoref{fig:asaga} shows experiment results for SAGA and ASAGA. The communication pattern in ASAGA is different from  ASGD because of the broadcast required to compute historical gradients. In ASAGA, the straggler and its delay intensity only affects the computation time of a worker and does not change the communication cost.  Therefore, the delay intensity does not have a linear effect on the overall runtime. However,  \autoref{fig:asaga} shows that increasing the delay intensity negatively affects the convergence rate of SAGA  while the ASAGA algorithm maintains the same convergence rate for different delay intensities. 

\begin{figure}
  \includegraphics[width=0.87\linewidth]{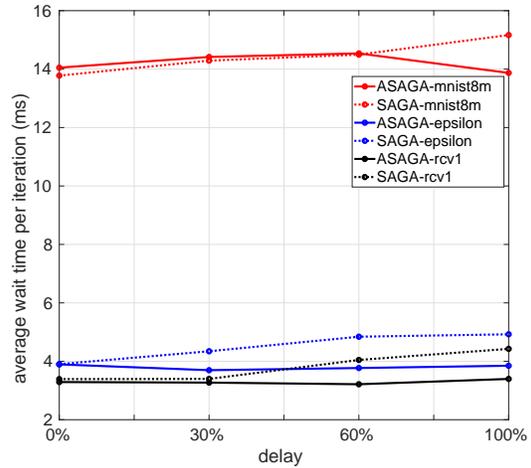}
  \caption{Average wait time per iteration with 8 workers for ASAGA and SAGA in ASYNC for different delay intensities.}
  \label{fig:asaga-wait-time}
\end{figure}



The workers' average  wait time for ASAGA  is shown in \autoref{fig:asaga-wait-time}. With an increase in delay intensity, workers in SAGA wait more for new tasks. The difference between the average wait time of SAGA and ASAGA is more noticeable when the delay increases to 100\%. In this case, the computation time is significant enough to affect the performance of the synchronous algorithm, however,  ASAGA has the same wait time for all delay intensities.

\mg{\textit{Production Cluster Stragglers:} 
Our PCS experiments are conducted on 32 workers with straggler patterns in real production clusters \cite{moreno2014analysis,ananthanarayanan2010reining}; these clusters are used frequently by machine learning practitioners. We use the straggler behaviors reported in previous research \cite{garraghan2016straggler,ouyang2016straggler} 
all of which are based on empirical analysis of production clusters from Microsoft Bing \cite{ananthanarayanan2010reining} and Google \cite{moreno2014analysis}. Empirical analysis from production clusters concluded that approximately 25\% of machines in cloud clusters are stragglers. From those, 80\% have a uniform probability of being delayed between 150\% to 250\% of average-task-completion time. The remaining 20\% of the stragglers have abnormal delays and are known as Long Tail workers. Long tail workers have a random delay between 250\% to 10$\times$.  From the 32 workers in our experiment, 6 are assigned a random delay between 150\%-250\% and two are long tail workers with a random delay over 250\% up to 10$\times$. The randomized delay seed is fixed across three executions of the same experiment.

}

\begin{figure}
  \includegraphics[width=0.87\linewidth]{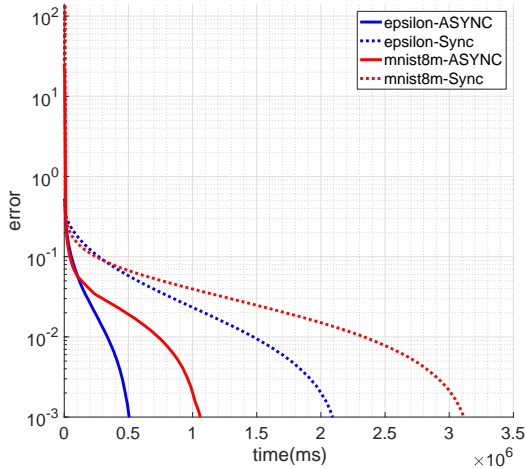}
  \caption{The performance of ASGD and SGD in ASYNC on 32 workers shown with ASYNC and SYNC respectively.}
  \label{fig:asgd-32}
\end{figure}

\begin{figure}
  \includegraphics[width=0.90\linewidth]{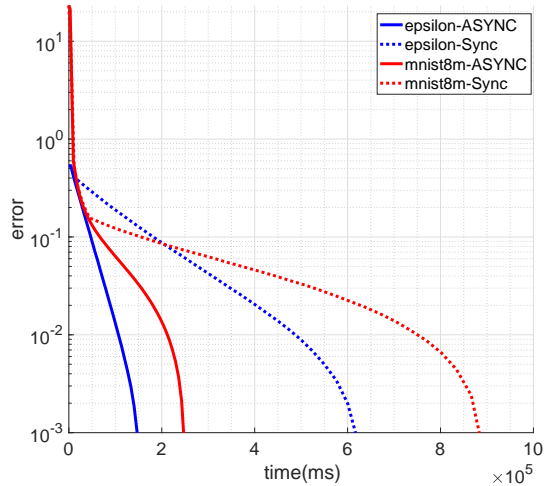}
  \caption{The performance of ASAGA and SAGA in ASYNC on 32 workers shown with ASYNC and SYNC respectively.}
  \label{fig:asaga-32}
\end{figure}

The performance of SGD and ASGD on 32 workers with PCS is shown in \autoref{fig:asgd-32}. As shown, ASGD converges to the solution considerably faster that SGD  and leads to a speedup of 3$\times$ for \textit{mnist8m} and 4$\times$ for \textit{epsilon}. 
From \autoref{fig:asaga-32}, ASAGA compared to SAGA obtains a speedup of 3.5$\times$ and 4$\times$ for \textit{mnist8m} and \textit{epsilon} respectively. The average wait time for both algorithms on 32 workers is shown in \autoref{tab:all-32}. The wait time increases considerably for all synchronous implementations which results in slower convergence of the synchronous methods.

\section{Related work}
To mitigate the negative effects of stale gradients on convergence, numerous optimization methods support asynchrony. The most widely used optimization algorithms with asynchrony are stochastic gradient methods \cite{dean2012large,recht2011hogwild}  and coordinate descent algorithms \cite{lian2015asynchronous}. Other work implement asynchrony by altering the execution  bound  staleness \cite{agarwal2011distributed,cipar2013solving}, by theoretically adapting the method to the stale gradients \cite{zhang2015staleness}, and by using barrier control strategies \cite{zhang2018adaptive,wang2017probabilistic}. 
Variance reduction approaches use the history of gradients to reduce the variance incurred by stochastic gradients and to improve convergence \cite{ming2018distributed,reddi2015variance,svrg}. Numerous algorithms implement variance reduction techniques in asynchronous methods, some of which include  ASAGA and DisSVRG \cite{ming2018distributed} which supports asynchrony in convex and non-convex  problems. 

The demand for large-scale machine learning has led to the development of numerous cloud and distributed computing frameworks. Commodity distributed dataflow systems such as Hadoop \cite{hadoop2011apache} and Spark \cite{zaharia2012resilient}, as well as libraries implemented on top of them such as Mllib\cite{meng2016mllib}, are optimized for coarse-grained, often bulk synchronous, parallel data transformations and thus do not provide asynchrony in their execution models
\cite{zaharia2012resilient,hadoop2011apache,mahout2008scalable,saadon2017iihadoop}. Recent work has modified frameworks such as Spark to support  asynchronous optimization methods.  ASIP \cite{gonzalez2015asynchronous} introduces a communication layer to Spark to support asynchrony, however, it only implements the  asynchronous parallel consistency model\cite{xing2015petuum} and does not support  history. Glint \cite{jagerman2016web}  integrates the parameter server model on Spark. However, it is designed for topic models with a specialized consistency model. 

Parameter server architectures such as \cite{xing2015petuum,qiao2018litz}  are widely used in distributed machine leaning since they support asynchrony in their execution models using a static dependency graph. Petuum \cite{xing2015petuum} implements the SSP execution model. Other parameter server frameworks include MLNET \cite{mai2015optimizing} and Litz \cite{qiao2018litz}. MLNET deploys a communication layer that uses tree-based overlays to implement distributed aggregation  to only communicate the aggregated updates without the support for individual communication of worker-results. These implementations do not support custom consistency models required by asynchronous optimization methods nor the history of gradients. Finally, numerous distributed computing frameworks have been developed to support specific applications. For example DistBelief \cite{dean2012large} and TensorFlow \cite{abadi2016tensorflow} support deep learning applications while fine-grained data processing systems such as RAY\cite{moritz2018ray} and Flink\cite{carbone2015apache} are designed for streaming problems. The frameworks can not be naturally extended to support  mini-batch optimization methods that require coarse-grained computation models.  

\section{Conclusion}
This work introduces the \framework{} framework that facilities the implementation of asynchrony and history in machine learning methods on cloud and distributed  platforms.  
Along with bookkeeping structures, the modules in \framework{} facilitate the implementation of numerous consistency models and history. 
\framework{} is built on top of Spark to benefit from Spark's in-memory computation model and fault tolerant execution. We present the programming model and interface that comes with \framework{} and implement the synchronous and asynchronous variants of two well-known optimization methods as examples. 
These examples only scratch the surface of the types of algorithms that can be implemented in \framework{}. We hope that \framework{} helps machine learning practitioners with the implementation and investigation to the promise of asynchronous optimization methods. 

\bibliography{sample-base.bib}{}
\bibliographystyle{IEEEtran}

\end{document}